\documentclass{article} 

\usepackage{amsmath}
\usepackage{bm}  
\usepackage{amssymb}

\newcommand{\Wfit}{M}
\usepackage{booktabs}

\usepackage{amsmath,amsthm,amsfonts}
\usepackage{graphicx}
\usepackage{booktabs}
\usepackage{parskip}
\usepackage[numbers]{natbib}
\newtheorem{mydef}{Definition}[section]
\newtheorem{thm}{Theorem}[section]

\renewenvironment{abstract}{\vskip.075in\centerline{\large\bf
Abstract}\vspace{0.5ex}\begin{quote}}{\par\end{quote}\vskip 1ex}

\title{Copula Processes}

\author{
Andrew Gordon Wilson\\
Zoubin Ghahramani \\
Department of Engineering \\
University of Cambridge \\
Cambridge, UK CB2 1PZ \\
\texttt{agw38@cam.ac.uk} \\
\texttt{zoubin@eng.cam.ac.uk}\\
}

\date{}

\begin{document}

\maketitle

\begin{abstract}
\normalsize
We define a copula process which describes the dependencies between arbitrarily many random variables
independently of their marginal distributions.  As an example, we develop a stochastic volatility model, Gaussian Copula 
Process Volatility (GCPV), to predict the latent standard deviations of a sequence of random variables.
To make predictions we use Bayesian inference, with the Laplace approximation, and with 
Markov chain Monte Carlo as an alternative.  We find both methods comparable.  We also find our model 
can outperform GARCH on simulated and financial data.  And unlike GARCH, GCPV can easily handle missing data, 
incorporate covariates other than time, and model a rich class of covariance structures.
\end{abstract}

\section{Introduction}
Imagine measuring the distance of a rocket as it leaves Earth, and wanting to know
how these measurements correlate with one another.  How much does the value of the
measurement at fifteen minutes depend on the measurement at five minutes? Once
we've learned this correlation structure, suppose we want to compare it to the
dependence between measurements of the rocket's velocity.  To do this, it is convenient to separate dependence from 
the marginal distributions of our measurements.  At any given time, a rocket's distance from Earth could have a Gamma distribution, while its velocity
could have a Gaussian distribution.  And separating dependence from marginal distributions is precisely what a copula function does.

While copulas have recently become popular, especially in financial applications \citep{Embrechts, li00},
as \citet{nelsen06} writes, ``the study of copulas and the role they play in probability, 
statistics, and stochastic processes is a subject still in its infancy.  There are many open 
problems\dots'' Typically only bivariate (and recently trivariate) copulas are being used and studied.  
In our introductory example, we are interested in learning
the correlations in different stochastic processes, and comparing them.  It would therefore
be useful to have a \textit{copula process}, which can describe the dependencies
between arbitrarily many random variables independently of their marginal distributions. We define such a process.
As an example, we develop a stochastic volatility model, \textit{Gaussian Copula Process Volatility} (GCPV).
In doing so, we provide a Bayesian framework for the learning the marginal distributions
and dependency structure of what we call a \textit{Gaussian copula process}.

The \textit{volatility} of a random variable is its standard deviation.  Stochastic volatility models
are used to predict the volatilities in a \textit{heteroscedastic} sequence -- a sequence of
random variables with different variances, like distance measurements of a rocket
as it leaves the Earth.  As the rocket gets further away, the variance on the measurements
increases.  Heteroscedasticity is especially important in econometrics; the returns
on equity indices, like the S\&P 500, or on currency exchanges, are heteroscedastic. 
Indeed, in 2003, Robert Engle won the Nobel Prize in economics 
``for methods of analyzing economic time series with time-varying volatility''.  
GARCH \citep{bollerslev86}, a generalized version of Engle's ARCH, is arguably unsurpassed for predicting the volatility of returns on
equity indices and currency exchanges \citep{granger05, hansen05, engle2009}.  GCPV can outperform GARCH,
and is competitive on financial data that especially suits GARCH \citep{bollerslev96, mccullough98, brooks2001}.
Before introducing GCPV, we first discuss copulas and then introduce our copula process.  For a review of Gaussian processes, see \citet{rasmussen06}.

\section{Copulas}
Copulas are important because they separate the dependency structure between random variables from their marginal distributions.
Intuitively, we can describe the dependency structure of any multivariate joint distribution $H(x_1,\dots,x_n)=P(X_1 \leq x_1, \dots X_n \leq x_n)$
through a two step process.  First we take each univariate random variable $X_i$ and transform it through its cumulative 
distribution function (cdf) $F_i$ to get $U_i = F_i(X_i)$, a uniform random variable.  We then express the dependencies between these transformed variables 
through the $n$-copula $C(u_1,\dots,u_n)$.  Formally, an $n$-copula $C: [0,1]^n \to [0,1]$ is a multivariate cdf with 
uniform univariate marginals: $C(u_1,u_2,\dots,u_n) = P(U_1 \leq u_1, U_2 \leq u_2, \dots, U_n \leq u_n)$, 
where $U_1,U_2,\dots,U_n$ are standard uniform random variables.  \citet{sklar59} precisely expressed our intuition
in the theorem below.
\newline
\begin{thm}  
Sklar's theorem \\
Let H be an n-dimensional distribution function with marginal distribution functions $F_1,F_2,\dots,F_n$.  
Then there exists an $n$-copula C such that for all 
$(x_1,x_2,\dots,x_n) \in [-\infty,\infty]^n$, 
\begin{equation}
H(x_1,x_2,\dots,x_n) = C(F_1(x_1),F_2(x_2),\dots,F_n(x_n)) = C(u_1,u_2,\dots,u_n).
\end{equation}
If $F_1,F_2,\dots,F_n$ are all continuous then C is unique; otherwise C is uniquely determined on 
$\operatorname{Range}{F_1}\times \operatorname{Range}{F_2} \times \dots \times \operatorname{Range}{F_n}$.  Conversely, if 
C is an $n$-copula and $F_1,F_2,\dots,F_n$ are distribution functions, then the function H is an n-dimensional
distribution function with marginal distribution functions $F_1,F_2,\dots,F_n$.
\end{thm}

As a corollary, if $F_i^{(-1)}(u) = \inf\{x: F(x) \geq u\}$, the \textit{quasi-inverse} of $F_i$, then for all $u_1,u_2,\dots,u_n \in [0,1]^n$,
\begin{equation}
 C(u_1,u_2,\dots,u_n) = H(F_1^{(-1)}(u_1),F_2^{(-1)}(u_2),\dots,F_n^{(-1)}(u_n)).
 \label{eqn: construct} 
\end{equation}

In other words, \eqref{eqn: construct} can be used to construct a copula.  For example, the bivariate Gaussian copula is defined as
\begin{equation}
 C(u,v) = \Phi_{\rho}(\Phi^{-1}(u),\Phi^{-1}(v)),
 \label{eqn: gausscopula}
\end{equation}
where $\Phi_{\rho}$ is a bivariate Gaussian cdf with correlation coefficient $\rho$, and $\Phi$ is the standard
univariate Gaussian cdf.  \citet{li00} popularised the bivariate Gaussian copula, by showing how it could be used to 
study financial risk and default correlation, using credit derivatives as an example.  

By substituting $F(x)$ for $u$ and $G(y)$ for $v$ in equation \eqref{eqn: gausscopula}, we have a bivariate distribution
$H(x,y)$, with a Gaussian dependency structure, and marginals $F$ and $G$.  \,Regardless of $F$ and $G$,  
the resulting $H(x,y)$ can still be uniquely expressed as a Gaussian copula, so long as $F$ and $G$ are continuous.  It is then a
copula itself that captures the underlying dependencies between random variables, regardless of their marginal distributions.
For this reason, copulas have been called \textit{dependence functions} \citep{deheuvals78,kimeldorf82}.  
\citet{nelsen06} contains an extensive discussion of copulas.

\section{Copula Processes}
Imagine choosing a covariance function, and then drawing a sample function at some finite number of points
from a Gaussian process.  The result is a sample from a collection of 
Gaussian random variables, with a dependency structure encoded by the specified covariance function.  
Now, suppose we transform each of these values through a univariate Gaussian cdf, such that we have a sample 
from a collection of uniform random variables.  These uniform random variables also have this underlying Gaussian process dependency structure.  
One might call the resulting values a draw from a \textit{Gaussian-Uniform process}.  We could subsequently
put these values through an inverse beta cdf, to obtain a draw from what could be called a 
\textit{Gaussian-Beta process}: the values would be a sample from beta random variables, again with
an underlying Gaussian process dependency structure.  Alternatively, we could transform the uniform values
with different inverse cdfs, which would give a sample from different random variables, with 
dependencies encoded by the Gaussian process.

The above procedure is a means to generate samples from arbitrarily many random variables, with 
arbitrary \textit{marginal} distributions, and desired dependencies.  It is an example of how to use what we call a \textit{copula process} -- 
in this case, a \textit{Gaussian copula process}, since a Gaussian copula describes 
the underlying dependency structure of a finite number of samples.  We can now formally define a copula process.
\newline
\begin{mydef} Copula Process \newline
Let $\{W_t\}$ be a collection of random variables indexed by $t \in \mathcal{T}$, with marginal distribution functions
$F_{t}$, and let $Q_t = F_t(W_t)$.  Further, let $\mu$ be a stochastic process measure with marginal distribution
functions $G_t$, and joint distribution function $H$.  Then $W_t$ is copula process distributed with base measure
$\mu$, or $W_t \sim \mathcal{CP}(\mu)$, if and only if for all $n \in \mathbb{N}$, $a_i \in \mathbb{R}$,
\begin{equation}
 P(\bigcap_{i=1}^n \{G_{t_i}^{(-1)}(Q_{t_i}) \leq a_i\}) = H_{t_1,t_2,\dots,t_n}(a_1,a_2,\dots,a_n).
\end{equation}
\end{mydef}
Note that each $Q_{t_i} \sim \operatorname{Uniform}(0,1)$, and that $G_{t_i}^{(-1)}$ is the quasi-inverse of $G_{t_i}$,
as it was previously defined.\newline
\begin{mydef} Gaussian Copula Process \newline
$W_t$ is Gaussian copula process distributed if it is copula process distributed and the base measure $\mu$ is a Gaussian process.
If there is a mapping $\Psi$ such that $\Psi(W_t) \sim \mathcal{GP}(m(t),k(t,t'))$, then we write $W_t \sim \mathcal{GCP}(\Psi,m(t),k(t,t'))$.
\end{mydef}

For example, if we have $W_t \sim \mathcal{GCP}$ with $m(t)=0$ and $k(t,t) = 1$, then in our definition of a copula process, $G_t = \Phi$, 
the standard univariate Gaussian cdf, and $H$ is the usual GP joint distribution function.  Supposing this GCP is a \textit{Gaussian-Beta process},
then $\Psi = \Phi^{-1} \circ F_B$, where $F_B$ is a univariate Beta cdf.  One could similarly define other copula processes.

We described generally how a copula process can be used to generate samples of arbitrarily many random variables with desired marginals 
and dependencies.  We now develop a specific and practical application of this framework.  We introduce a stochastic volatility model, 
\textit{Gaussian Copula Process Volatility} (GCPV), as an example of how to learn the joint distribution of arbitrarily many random variables, 
the marginals of these random variables, and to make predictions.  To do this, we fit a Gaussian copula process by using a type of 
Warped Gaussian Process \citep{snelson03}.  However, our methodology varies substantially from \citet{snelson03}, since we are doing inference 
on latent variables as opposed to observations, which is a much greater undertaking that involves approximations, and we are doing so in a different
context. 

\section{Gaussian Copula Process Volatility}
Assume we have a sequence of observations $\bm{y}=(y_1,\dots,y_n)^{\top}$ at times $\bm{t}=(t_1,\dots,t_n)^{\top}$.  The observations are random variables with 
different latent standard deviations.  We therefore have $n$ unobserved standard deviations, $\sigma_1,\dots,\sigma_n$, and want to  
learn the correlation structure between these standard deviations, and also to predict the distribution of  $\sigma_*$ at some unrealised time $t_*$. 

We model the standard deviation function as a Gaussian copula process:
\begin{equation}
\sigma_t \sim \mathcal{GCP}(g^{-1},0,k(t,t')).
\label{eqn: gcpv}
\end{equation}
Specifically, 
\begin{align}
f \sim \mathcal{GP}(m(t)=0, k(t,t'))\label{eqn: gp}\\
\sigma(t) = g(f(t),\bm{\omega}) \label{eqn: gpalt} \\
y|t \sim \mathcal{N}(0,\sigma^2(t)),
\end{align}
where $g$ is a monotonic warping function, parametrized by $\bm{\omega}$.  For each of the observations $\bm{y}=(y_1,\dots,y_n)^{\top}$ we
have corresponding GP latent function values $\bm{f}=(f_1,\dots,f_n)^{\top}$, where $\sigma(t_i) = g(f_i,\bm{\omega})$, using the 
shorthand $f_i$ to mean $f(t_i)$. 

$\sigma_t \sim \mathcal{GCP}$, because any finite sequence $(\sigma_1,\dots, \sigma_p)$ is distributed as a Gaussian copula:
\begin{align}
P(\sigma_1 \leq a_1, \dots, \sigma_p \leq a_p) = P(g^{-1}(\sigma_1) \leq g^{-1}(a_1), \dots, g^{-1}(\sigma_p) \leq g^{-1}(a_p))  \label{eqn: gcpsigma} \\
= \Phi_{\Gamma}(g^{-1}(a_1), \dots, g^{-1}(a_p)) = \Phi_{\Gamma}(\Phi^{-1}(F(a_1)),\dots,\Phi^{-1}(F(a_p))) \notag \\
= \Phi_{\Gamma}(\Phi^{-1}(u_1),\dots,\Phi^{-1}(u_p)) = C(u_1,\dots,u_p), \notag
\end{align}
where $\Phi$ is the standard univariate Gaussian cdf (supposing $k(t,t)=1$), 
$\Phi_{\Gamma}$ is a multivariate Gaussian cdf with covariance matrix $\Gamma_{ij} = \textrm{cov}(g^{-1}(\sigma_i),g^{-1}(\sigma_j))$,
and $F$ is the marginal distribution of each $\sigma_i$.  In \eqref{eqn: gcpv}, we have $\Psi = g^{-1}$, because it is $g^{-1}$
which maps $\sigma_t$ to a GP.\, The specification in \eqref{eqn: gcpv} is equivalently expressed by \eqref{eqn: gp} and \eqref{eqn: gpalt}.  
With GCPV, the form of $g$ is learned so that $g^{-1}(\sigma_t)$ is best modelled by a GP.  By learning $g$, we learn the marginal of each 
$\sigma$: $F(a) = \Phi(g^{-1}(a))$ for $a \in \mathbb{R}$.  Recently, a different sort of `kernel copula process' has been used, 
where the marginals of the variables being modelled are not learned \citep{jaimungal09}.  Further, we also consider a more subtle and flexible form of our model, where the function 
$g$ itself is indexed by time: $g = g_t(f(t),\bm{\omega})$.  We only assume that the marginal distributions of $\sigma_t$ are stationary over `small' time periods, 
and for each of these time periods (5)-(7) hold true.  We return to this in the final discussion section.

Here we have assumed that each observation, conditioned on knowing its variance, is normally distributed with zero mean.
This is a common assumption in heteroscedastic models.  The zero mean and normality assumptions can be relaxed 
and are not central to this paper.
  
\section{Predictions with GCPV}  
Ultimately, we wish to infer $p(\sigma(t_*)|\bm{y},\bm{z})$, where $\bm{z} = \{\bm{\theta},\bm{\omega}\}$, and
$\bm{\theta}$ are the hyperparameters of the GP covariance function.  To do this, we sample from
\begin{equation}
p(f_*|\bm{y},\bm{z})= \int p(f_*|\bm{f},\bm{\theta})p(\bm{f}|\bm{y},\bm{z}) d\bm{f}
\label{eqn: fstary}
\end{equation}
and then transform these samples by $g$.  Letting $(C_{\bm{f}})_{ij}$ = $\delta_{ij} g(f_i,\bm{\omega})^2$, where $\delta_{ij}$ is the Kronecker delta,
$K_{ij} = k(t_i,t_j)$, $(\bm{k}_*)_i = k(t_*,t_i)$, we have
\begin{align} 
p(\bm{f}|\bm{y},\bm{z}) = \mathcal{N}(\bm{f}; 0, K) \mathcal{N}(\bm{y}; 0, C_{\bm{f}}) / p(\bm{y}|\bm{z}), \label{eqn: fy} \\
p(f_*|\bm{f},\bm{\theta})= \mathcal{N}(\bm{k}_*^{\top}K^{-1}\bm{f},k(t_*,t_*)-\bm{k}_*^{\top}K^{-1}\bm{k}_*).
\end{align}
We also wish to learn $\bm{z}$, which we can do by finding the $\bm{\hat{z}}$ that maximizes the marginal likelihood,
\begin{equation}
p(\bm{y}|\bm{z}) = \int p(\bm{y}|\bm{f},\bm{\omega})p(\bm{f}|\bm{\theta}) d\bm{f}.
\label{eqn: ml} 
\end{equation}
Unfortunately, for many functions $g$, \eqref{eqn: fstary} and \eqref{eqn: ml} are intractable.
Our methods of dealing with this can be used in very general circumstances, where one has a Gaussian process prior, 
but an (optionally parametrized) non-Gaussian likelihood.  We use the Laplace approximation to estimate $p(\bm{f}|\bm{y},\bm{z})$ as a Gaussian.
Then we can integrate \eqref{eqn: fstary} for a Gaussian approximation to $p(f_*|\bm{y},\bm{z})$, which we sample from to 
make predictions of $\sigma_*$.  Using Laplace, we can also find an expression for
an approximate marginal likelihood, which we maximize to determine $\bm{z}$.  While we 
always use Laplace to determine $\bm{z}$, we compare to a full Laplace solution by also using
Markov chain Monte Carlo to sample from $p(f_*|\bm{y},\bm{z})$. 

Let us now relate the above to the Gaussian copula in \eqref{eqn: gcpsigma}.
The prior $\Gamma_{ij} = \textrm{cov}(g^{-1}(\sigma_i),g^{-1}(\sigma_j)) = \textrm{cov}(f_i,f_j) = k(t_i,t_j)$.
The posterior $\Gamma_{ij}$ can be estimated as the covariance matrix of the Laplace approximation for
$p(\bm{f}|\bm{y})$. Also, since each component of $\bm{f}$ is transformed separately, such that $\sigma(t_i)=g(f(t_i))$, we have
\begin{equation}
p(\bm{\sigma}|\bm{y},\bm{z}) = \displaystyle \big{[}\prod_{i=1}^{N} \frac{df_i}{d\sigma_i}\big{]}p(\bm{f}|\bm{y},\bm{z}) = \displaystyle \big{[}\prod_{i=1}^{N} \frac{1}{g'(f_i,\bm{\omega})}\big{]} p(\bm{f}|\bm{y},\bm{z}).
\end{equation}
One can use this to simulate from the joint distribution over the deviations.

\subsection{Laplace Approximation}
The goal is to approximate \eqref{eqn: fy} with a Gaussian, so that we can evaluate \eqref{eqn: fstary} and
\eqref{eqn: ml} and make predictions.  In doing so, we follow \citet{rasmussen06} in their treatment of 
Gaussian process classification, except we use a parametrized likelihood, and modify Newton's method.

First, consider as an objective function the logarithm of an unnormalized \eqref{eqn: fy}:
\begin{equation}
s(\bm{f}|\bm{y},\bm{z}) = \log p(\bm{y}|\bm{f},\bm{\omega}) + \log p(\bm{f}|\bm{\theta}). 
\label{eqn: obj} 
\end{equation}
The Laplace approximation uses a second order Taylor expansion about the $\bm{\hat{f}}$ 
which maximizes \eqref{eqn: obj}, to find an approximate objective $\tilde{s}(\bm{f}|\bm{y},\bm{z})$.
So the first step is to find $\bm{\hat{f}}$, for which we use Newton's method.  The Newton update 
is $\bm{f}^{\text{new}} = \bm{f} - (\nabla\nabla s(\bm{f}))^{-1}\nabla s(\bm{f})$.
Differentiating \eqref{eqn: obj}, 
\begin{align}
\nabla s(\bm{f}|\bm{y},\bm{z}) = \nabla \log p(\bm{y}|\bm{f},\bm{\omega}) - K^{-1}\bm{f} \\
\nabla \nabla s(\bm{f}|\bm{y},\bm{z}) = \nabla \nabla \log p(\bm{y}|\bm{f},\bm{\omega}) - K^{-1} = -W - K^{-1},
\end{align}
where $W$ is the diagonal matrix $-\nabla \nabla \log p(\bm{y}|\bm{f},\bm{\omega})$. 

If the likelihood function $p(\bm{y}|\bm{f},\bm{\omega})$ is not log concave, then $W$ may have negative entries.  \citet{vanhatalo09} found this to be problematic when
doing Gaussian process regression with a Student-t likelihood.  They instead use an expectation-maximization (EM) algorithm for finding $\bm{\hat{f}}$, and iterate ordered rank one Cholesky 
updates to evaluate the Laplace approximate marginal likelihood.  But EM can converge slowly, especially near a local optimum,
and each of the rank one updates is vulnerable to numerical instability. 
With a small modification of Newton's method, we often get close to quadratic convergence for finding $\bm{\hat{f}}$, and can 
evaluate the Laplace approximate marginal likelihood in a numerically stable fashion, with no approximate Cholesky factors, and optimal computational requirements.

At a maximum, the negative Hessian of the objective function, $W+K^{-1}$, is positive definite.  On each iteration of Newton's method, we form $\Wfit$ by 
setting all negative entries of $W$ to zero.  Since $K^{-1}$ is positive definite, and the eigenvalues of $\Wfit+K^{-1}$ are greater than or equal to 
the eigenvalues of $K^{-1}$, $\Wfit+K^{-1}$ is always positive definite.  Using $\Wfit$ in place of $W$ decreases the 
Newton step size, and changes the direction of steps.  We are always stepping towards a local maximum, and will converge, barring rare pathologies.

Furthermore, we reformulate our optimization in terms of $B = I + \Wfit^{\frac{1}{2}}K\Wfit^{\frac{1}{2}}$, which is often well 
conditioned: it has eigenvalues no smaller than 1, and no larger than $1+n\text{ }\textrm{max}_{ij}(K_{ij})/4$.  
Letting $Q = \Wfit^{\frac{1}{2}}B^{-1}\Wfit^{\frac{1}{2}}$, we find
$(K^{-1}+\Wfit)^{-1} = K - KQK$, and the Newton update becomes $\bm{f}^{new} = K\bm{a}$, where 
$\bm{a} = \bm{b} - QK\bm{b}$, and $\bm{b} = \Wfit\bm{f} + \nabla \log p(\bm{y}|\bm{f})$.  
With these Newton updates we find $\bm{\hat{f}}$ and get an expression for $\tilde{s}$ which we use to approximate \eqref{eqn: fy} and \eqref{eqn: ml}.

The approximate marginal likelihood is given by $\int \exp(\tilde{s}) d\bm{f}$.  Taking its logarithm,
\begin{equation}
\log q(\bm{y}|\bm{z}) = -\frac{1}{2}\bm{\hat{f}}^{\top}\bm{a}_{\bm{\hat{f}}} + \log p(\bm{y}|\bm{\hat{f}}) - \frac{1}{2}\log |B_{\bm{\hat{f}}}|,
\label{eqn: aml}
\end{equation}
where $B_{\bm{\hat{f}}}$ is $B$ evaluated at $\bm{\hat{f}}$, and $\bm{a}_{\bm{\hat{f}}}$ is a numerically stable evaluation of $K^{-1}\bm{\hat{f}}$.

To learn the parameters $\bm{z}$, we use conjugate gradient descent to maximize \eqref{eqn: aml} with respect to $\bm{z}$.  Since $\bm{\hat{f}}$ is a function of $\bm{z}$,
we initialize $\bm{z}$, and update $\bm{\hat{f}}$ every time we vary $\bm{z}$.  Once we have found an optimum $\bm{\hat{z}}$, we can make predictions.  By exponentiating 
$\tilde{s}$, we find a Gaussian approximation to the posterior \eqref{eqn: fy}, 
$q(\bm{f}|\bm{y},\bm{z}) = \mathcal{N}(\bm{\hat{f}},K-KQK$). The product
of this approximate posterior with $p(f_*|\bm{f})$ is Gaussian.  Integrating this product, we approximate $p(f_*|\bm{y},\bm{z})$ as 
\begin{equation}
q(f_*|\bm{y},\bm{z}) = \mathcal{N}(\bm{k}_*^{\top}\nabla \log p(\bm{y}|\bm{\hat{f}}), k(t_*,t_*)-\bm{k}_*^{\top}Q\bm{k_*}).
\label{eqn: afstar}
\end{equation}
Given $n$ training observations, the cost of each Newton iteration is dominated by computing
$chol(B)$, which takes $\mathcal{O}(n^3)$ operations.  The objective function typically changes by less than $10^{-6}$ after 3 iterations. 
Once Newton's method has converged, it takes only $\mathcal{O}(1)$ operations to draw from $q(f_*|\bm{y},\bm{z})$ and make predictions.

\subsection{Markov chain Monte Carlo}
We use Markov chain Monte Carlo (MCMC) to sample from \eqref{eqn: fy}, so that we can later sample from $p(\sigma_*|\bm{y},\bm{z})$ 
to make predictions.  Sampling from \eqref{eqn: fy} is difficult, because
the variables $\bm{f}$ are strongly coupled by a Gaussian process prior.  We use a new technique, Elliptical Slice Sampling \citep{murray10}, and find it 
extremely effective for this purpose.  It was specifically designed to sample from posteriors with correlated Gaussian priors.  It has no free parameters, 
and jointly updates every element of $\bm{f}$.  For our setting, it is over 100 times as fast as axis aligned slice sampling with univariate updates.

To make predictions, we take $J$ samples of $p(\bm{f}|\bm{y},\bm{z})$, \{$\bm{f}^{1},\dots,\bm{f}^{J}$\}, and then approximate \eqref{eqn: fstary} as a mixture 
of $J$ Gaussians:
\begin{equation}
p(f_*|\bm{y},\bm{z}) \approx \displaystyle \frac{1}{J} \sum_{i=1}^{J} p(f_*|\bm{f}^{i},\bm{\theta}).
\label{eqn: afysamp}
\end{equation}
Each of the Gaussians in this mixture have equal weight.  So for each sample of $f_*|\bm{y}$, we uniformly choose a random $p(f_*|\bm{f}^{i},\bm{\theta})$ and draw a sample.
In the limit $J \to \infty$, we are sampling from the exact $p(f_*|\bm{y},\bm{z})$.  Mapping these samples through $g$ gives samples from $p(\sigma_*|\bm{y},\bm{z})$.

After one $\mathcal{O}(n^3)$ and one $\mathcal{O}(J)$ operation, a draw from \eqref{eqn: afysamp} takes $\mathcal{O}(1)$ operations.

\subsection{Warping Function}
The warping function, $g$, maps $f_i$, a GP function value, to $\sigma_i$, a standard deviation.  Since $f_i$ can take any value in $\mathbb{R}$, and $\sigma_i$ can
take any non-negative real value, $g\!: \mathbb{R} \to \mathbb{R}^{+}$.  For each $f_i$ to correspond to a unique deviation, $g$ must also be one-to-one.  We use 
\begin{equation}
g(x,\bm{\omega})= \displaystyle \sum_{j=1}^{K} a_j\log[\exp[b_j(x+c_j)] + 1],\quad a_j,b_j > 0.
\label{eqn: warping}
\end{equation}
This is monotonic, positive, infinitely differentiable, asymptotic towards zero as $x \to -\infty$, and tends to $(\sum_{j=1}^K a_jb_j)x$ as $x \to \infty$.
In practice, it is useful to add a small constant to \eqref{eqn: warping}, to avoid rare situations
where the parameters $\bm{\omega}$ are trained to make $g$ extremely small for certain inputs, at the expense of a good overall fit; this can happen when the
parameters $\bm{\omega}$ are learned by optimizing a likelihood.  A suitable constant could be one tenth the absolute value of the smallest nonzero observation.

By inferring the parameters of the warping function, or distributions of these parameters, we are learning a transformation
which will best model $\sigma_t$ with a Gaussian process.  The more flexible the warping function, the more potential there is to improve 
the GCPV fit -- in other words, the better we can estimate the `perfect' transformation. To test the importance of this flexibility,
we also try a simple unparametrized warping function, $g(x) = e^{x}$.  In related work, \citet{goldberg98}
place a GP prior on the log noise level in a standard GP regression model on observations, except for inference 
they use Gibbs sampling, and a high level of `jitter' for conditioning.

Once $g$ is trained, we can infer the marginal distribution of each $\sigma$: $F(a) = \Phi(g^{-1}(a))$,
for $a \in \mathbb{R}$.  This suggests an alternate way to initialize $g$: we can initialize $F$ as a mixture of Gaussians, and then map
through $\Phi^{-1}$ to find $g^{-1}$.  Since mixtures of Gaussian distributions are dense in the set of probability distributions, 
we could in principle find the `perfect' $g$ using an infinite mixture of Gaussians \citep{rasmussen00}. 

\section{Experiments}
In our experiments, we predict the latent standard deviations $\bm{\sigma}$ of observations $\bm{y}$ at times $\bm{t}$, and also $\bm{\sigma_*}$ at 
unobserved times $\bm{t_*}$.  To do this, we use two versions of GCPV.\, The first variant, which we simply refer to as GCPV, uses the
warping function \eqref{eqn: warping} with $K=1$, and squared exponential covariance function, $k(t,t') = A\exp(-(t-t')^2/l^2)$, with $A=1$.
The second variant, which we call GP-EXP, uses the unparametrized warping function $e^{x}$, and the same covariance function, except the 
amplitude $A$ is a trained hyperparameter.  The other hyperparameter $l$ is called the \textit{lengthscale} of the covariance function.
The greater $l$, the greater the covariance between $\sigma_t$ and $\sigma_{t+a}$ for $a \in \mathbb{R}$.  We train hyperparameters by maximizing 
the Laplace approximate log marginal likelihood \eqref{eqn: aml}.

We then sample from $p(f_* | \bm{y})$ using the Laplace approximation \eqref{eqn: afstar}.  We also do this using 
MCMC \eqref{eqn: afysamp} with $J=10000$, after discarding a previous $10000$ samples of $p(\bm{f}|\bm{y})$ as burn-in.  
We pass these samples of $f_*|\bm{y}$ through $g$ and $g^2$ to draw from $p(\sigma_* | \bm{y})$ and $p(\sigma_*^2 | \bm{y})$, 
and compute the sample mean and variance of $\sigma_* | \bm{y}$.  We use the sample mean as a point predictor, and the sample 
variance for error bounds on these predictions, and we use $10000$ samples to compute these quantities.  For GCPV we use
Laplace and MCMC for inference, but for GP-EXP we only use Laplace.  We compare predictions to GARCH(1,1), which has been shown 
in extensive and recent reviews to be competitive with other GARCH variants, and more sophisticated models \citep{granger05, hansen05, engle2009}.  
We use the Matlab Econometrics Toolbox implementation of GARCH.

We make forecasts of volatility, and we predict historical volatility. By `historical volatility' we mean the volatility at
observed time points, or between these points.  Uncovering historical volatility is important.  It could, for instance,
be used to study what causes fluctuations in the stock market, or to understand physical systems. 

To evaluate our model, we use the Mean Squared Error (MSE) between the true variance, or proxy for the truth, and the
predicted variance.  Although likelihood has advantages, we are limited in space, and we wish to harmonize
with the econometrics literature, and other assessments of volatility models, where MSE is the standard.  In a 
similar assessment of volatility models, \citet{engle2009} found that MSE and quasi-likelihood rankings were comparable.

When the true variance is unknown we follow \citet{engle2009} and use squared observations as a proxy for the truth, to compare our 
model to GARCH.\footnote{Since each observation $y$ is assumed to have zero mean and variance $\sigma^2$, $\mathbb{E}[y^2] = \sigma^2$.}  
The more observations, the more reliable these performance estimates will be.  However, not many observations (e.g. 100) are needed for a 
stable ranking of competing models; in \citet{engle2009}, the rankings derived from high frequency squared observations are similar  
to those derived using daily squared observations.

\subsection{Simulations}
We simulate observations from $\mathcal{N}(0,\sigma^{2}(t))$, using $\sigma(t) = \sin(t)\cos(t^2) + 1$, at $\bm{t} = (0, 0.02, 0.04, \dots, 4)^{\top}$.
We call this data set \texttt{TRIG}.  We also simulate using a standard deviation that jumps from $0.1$ to $7$ and back, at times
$\bm{t} = (0, 0.1, 0.2, \dots, 6)^{\top}$.  We call this data set \texttt{JUMP}. To forecast, we use all observations up until the current time point, 
and make 1, 7, and 30 step ahead predictions.  So, for example, in \texttt{TRIG} we start by observing $t=0$, and make forecasts at $t=0.02, 0.14, 0.60$.
Then we observe $t=0, 0.02$ and make forecasts at $t=0.04, 0.16, 0.62$, and so on, until all data points have been observed.  For historical volatility, 
we predict the latent $\sigma_t$ at the observation times, which is safe since we are comparing to the true volatility, which is not used in 
training; the results are similar if we interpolate.  Figure 1 panels a) and b) show the true volatility for \texttt{TRIG} and \texttt{JUMP} respectively,
alongside GCPV Laplace, GCPV MCMC, GP-EXP Laplace, and GARCH(1,1) predictions of historical volatility.  
Table 1 shows the results for forecasting and historical volatility.  

In panel a) we see that GCPV more accurately captures the dependencies between $\sigma$ at different times points than GARCH: if we manually
decrease the lengthscale in the GCPV covariance function, we can replicate the erratic GARCH behaviour, which inaccurately suggests that the
covariance between $\sigma_t$ and $\sigma_{t+a}$ decreases quickly with increases in $a$.  We also see that GCPV with an
unparametrized exponential warping function tends to overestimates peaks and underestimate troughs.  In panel b), the volatility is extremely 
difficult to reconstruct or forecast -- with no warning it will immediately and dramatically increase or decrease.  This behaviour is not
suited to a smooth squared exponential covariance function. Nevertheless, GCPV outperforms GARCH, especially in regions of low volatility.  
We also see this in panel a) for $t \in (1.5,2)$.  GARCH is known to respond slowly to large returns, and to overpredict volatility \citep{tsay02}.  In
\texttt{JUMP}, the greater the peaks, and the smaller the troughs, the more GARCH suffers, while GCPV is mostly robust to these changes.

\subsection{Financial Data}
The returns on the daily exchange rate between the Deutschmark (DM) and the Great Britain Pound (GBP) from
1984 to 1992 have become a benchmark for assessing the performance of GARCH models \citep{bollerslev96, mccullough98, brooks2001}.  
This exchange data, which we refer to as \texttt{DMGBP}, can be obtained from \texttt{www.datastream.com}, and the returns are calculated 
as $r_t = \log(P_{t+1}/P_{t})$, where $P_t$ is the number of DM to GBP on day $t$.
The returns are assumed to have a zero mean function. 

We use a rolling window of the previous 120 days of returns to make 1, 7, and 30 day ahead volatility forecasts,
starting at the beginning of January 1988, and ending at the beginning of January 1992 (659 trading days).
Every 7 days, we retrain the parameters of GCPV and GARCH.  Every time we retrain parameters, 
we predict historical volatility over the past 120 days.  The average MSE for these historical predictions is given in Table 1,
although they should be observed with caution; unlike with the simulations, the \texttt{DMGBP} historical predictions are trained
using the same data they are assessed on.  In Figure 1c), we see that the GARCH one day ahead forecasts 
are lifted above the GCPV forecasts, but unlike in the simulations, they are now operating on a similar lengthscale. 
This suggests that GARCH could still be overpredicting volatility, but that GCPV has adapted its estimation of 
how $\sigma_t$ and $\sigma_{t+a}$ correlate with one another.  Since GARCH is suited to this financial data set, 
it is reassuring that GCPV predictions have a similar time varying structure.  Overall, GCPV and GARCH
are competitive with one another for forecasting currency exchange returns, as seen in Table 1.
Moreover, a learned warping function $g$ outperforms an unparametrized one, and a full Laplace solution is comparable to
using MCMC for inference, in accuracy and speed.  This is also true for the simulations. 
Therefore we recommend whichever is more convenient to implement.

\section{Discussion}
We defined a copula process, and as an example, developed a stochastic volatility model, GCPV,
which can outperform GARCH.\, With GCPV, the volatility $\sigma_t$ is distributed as a Gaussian Copula
Process, which separates the modelling of the dependencies between volatilities at different times
from their marginal distributions -- arguably the most useful property of a copula. Further,
GCPV fits the marginals in the Gaussian copula process by learning a warping function.  If we had 
simply chosen an unparametrized exponential warping function, we would incorrectly be assuming 
that the log volatilities are marginally Gaussian distributed.  Indeed, for the \texttt{DMGBP}
data, we trained the warping function $g$ over a 120 day period, and mapped its inverse 
through the univariate standard Gaussian cdf $\Phi$, and differenced, to estimate the marginal
probability density function (pdf) of $\sigma_t$ over this period.
The learned marginal pdf, shown in Figure 1d), is similar to a Gamma(4.15,0.00045) distribution.  However, in using a 
rolling window to retrain the parameters of $g$, we do not assume that the marginals of $\sigma_t$ are stationary;
we have a time changing warping function.  

While GARCH is successful, and its simplicity is attractive, our model is also
simple and has a number of advantages.  We can effortlessly handle missing data,
we can easily incorporate covariates other than time (like interest rates) in our covariance function,
and we can choose from a rich class of covariance functions -- squared exponential, Brownian motion,
Mat{\'e}rn, periodic, etc.  In fact, the volatility of high frequency intradaily returns on equity indices
and currency exchanges is cyclical \citep{andersen97}, and GCPV with a periodic covariance function is uniquely
well suited to this data.  And the parameters of GCPV, like the covariance function
lengthscale, or the learned warping function, provide insight into the underlying source of volatility, 
unlike the parameters of GARCH.  

Finally, copulas are rapidly becoming popular in applications, but often only bivariate copulas are being used.  
We introduced a copula process where one can learn the dependency structure between arbitrarily many random variables
independently of their marginal distributions.  We hope the Gaussian Copula Process Volatility model will encourage
other applications of copula processes.  More generally, we hope our work will help bring together the
machine learning and econometrics communities.

\begin{table}
\caption{MSE for predicting volatility.}
\small
\begin{center}
\begin{tabular}{l r r r r r}
\toprule
 Data set & Model & Historical & 1 step & 7 step & 30 step\\
\midrule
 \texttt{TRIG} & GCPV (LA) & $0.953 \times 10^{-1}$ & $0.588 \times 10^{0}$ & $0.951 \times 10^{0}$ & $1.71 \times 10^{0}$\\
 & GCPV (MCMC) & $0.760 \times 10^{-1}$ & $0.622 \times 10^{0}$ & $0.979 \times 10^{0}$ & $1.76 \times 10^{0}$\\
 & GP-EXP & $1.93 \times 10^{-1}$ & $0.646 \times 10^{0}$ & $1.36 \times 10^{0}$ & $1.15 \times 10^{0}$\\
 & GARCH & $9.38 \times 10^{-1}$ & $1.04 \times 10^{0}$ & $1.79 \times 10^{0}$ & $5.12 \times 10^{0}$\\
\midrule
 \texttt{JUMP} & GCPV (LA) & $0.588 \times 10^3$ & $0.891 \times 10^3$ & $1.38 \times 10^3$ & $0.135 \times 10^4$\\
 & GCPV (MCMC) & $1.21 \times 10^3$ & $0.951 \times 10^3$ & $1.37 \times 10^3$ & $0.135 \times 10^4$\\
 & GP-EXP & $1.43 \times 10^3$ & $1.76 \times 10^3$ & $6.95 \times 10^3$ & $1.47 \times 10^4$\\
 & GARCH & $1.88 \times 10^3$ & $1.58 \times 10^3$ & $3.43 \times 10^3$ & $0.565 \times 10^4$\\
\midrule
 \texttt{DMGBP} & GCPV (LA) & $2.43 \times 10^{-9}$ & $3.00 \times 10^{-9}$ & $3.08 \times 10^{-9}$ & $3.17 \times 10^{-9}$ \\
 & GCPV (MCMC) & $2.39 \times 10^{-9}$ & $3.00 \times 10^{-9}$ & $3.08 \times 10^{-9}$ & $3.17 \times 10^{-9}$\\
 & GP-EXP & $2.52 \times 10^{-9}$ & $3.20 \times 10^{-9}$ & $3.46 \times 10^{-9}$ & $5.14 \times 10^{-9}$ \\
 & GARCH & $2.83 \times 10^{-9}$ & $3.03 \times 10^{-9}$ & $3.12 \times 10^{-9}$ & $3.32 \times 10^{-9}$ \\
\bottomrule
\end{tabular}
\end{center}
\label{tab:gcpvtable}
\end{table}

\begin{figure}
\centering
\includegraphics[scale=.5]{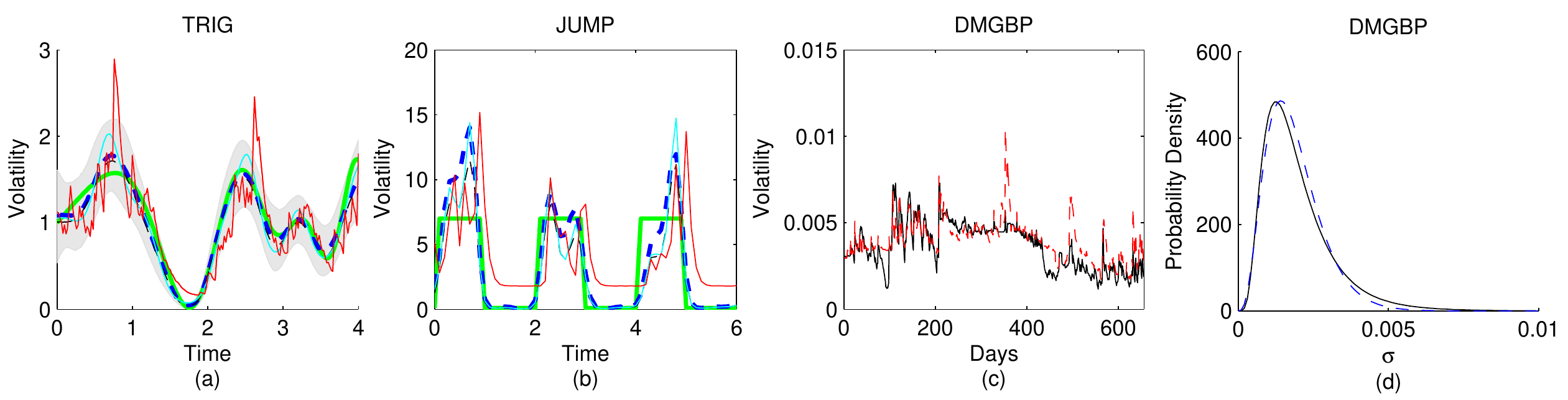}
\caption{\small Predicting volatility and learning its marginal pdf.  For a) and b), the true volatility, and
GCPV (MCMC), GCPV (LA), GP-EXP, and GARCH predictions, are shown respectively by a thick green line, 
a dashed thick blue line, a dashed black line, a cyan line, and a red line.  a) shows predictions of historical
volatility for \texttt{TRIG}, where the shade is a 95\% confidence interval about GCPV (MCMC) predictions. 
b) shows predictions of historical volatility for \texttt{JUMP}. In c), a black
line and a dashed red line respectively show GCPV (LA) and GARCH one day ahead volatility forecasts for \texttt{DMGBP}. In d), a
black line and a dashed blue line respectively show the GCPV learned marginal pdf of $\sigma_t$ in \texttt{DMGBP}
and a Gamma(4.15,0.00045) pdf.}
\label{fig:gcpvcompare}
\end{figure}

\textbf{Acknowledgments}: Thanks to Carl Edward Rasmussen and Ferenc Husz\'{a}r for helpful conversations.  AGW is supported
by an NSERC grant.

\bibliographystyle{unsrtnat}
\bibliography{mbib}

\end{document}